\documentclass[copyright,creativecommons]{eptcs}
\providecommand{\event}{E. Formenti (Ed.): AUTOMATA \& JAC 2012} 

\usepackage{graphicx}
\usepackage{wrapfloat}
\usepackage{makeidx}  

\usepackage{amsmath, amssymb}

\def\refname{Bibliography}

\title{A Simple Optimum-Time FSSP Algorithm for Multi-Dimensional Cellular Automata}
\def\titlerunning{An Optimum-Time FSSP Algorithm for Multi-Dimensional Cellular Automata}
\author{H{\normalsize IROSHI} U{\normalsize MEO}
\institute{University of Osaka Electro-Communication}
\email{umeo@cyt.osakac.ac.jp}
\and
K{\normalsize INUO} N{\normalsize ISHIDE}
\institute{University of Osaka Electro-Communication}
\email{nishide@cyt.osakac.ac.jp}
\and
K{\normalsize EISUKE} K{\normalsize UBO}
\institute{University of Osaka Electro-Communication}
\email{kubo@cyt.osakac.ac.jp}
}
\def\authorrunning{H. Umeo, K. Nishide \&  K. Kubo}

\begin{document} 

\maketitle

\begin{abstract}
The firing squad synchronization problem (FSSP) on cellular automata has been studied extensively for more than forty years, and a rich variety of synchronization algorithms have been proposed for not only one-dimensional arrays but two-dimensional arrays.  In the present paper, we propose a simple recursive-halving based optimum-time synchronization algorithm that can synchronize any rectangle arrays of size $m \times n$ with a general at one corner in $m + n + \max(m, n)-3 $ steps. The algorithm is a natural expansion of the well-known FSSP algorithms proposed by Balzer [1967], Gerken [1987], and Waksman [1966] and it can be easily expanded to three-dimensional arrays, even to multi-dimensional arrays with a general at any position of the array. The algorithm proposed is isotropic concerning the side-lengths of multi-dimensional arrays and its algorithmic correctness is transparent and easily verified.
\end{abstract}

\section{Introduction}

We study a synchronization problem that gives a finite-state protocol for synchronizing large-scale cellular automata.  The synchronization in cellular automata has been known as a firing squad synchronization problem (FSSP) since its development, in which it was originally proposed by J. Myhill in Moore [1964] to synchronize all parts of self-reproducing cellular automata. The problem has been studied extensively for more than forty years [1-23], and a rich variety of synchronization algorithms have been proposed for not only one-dimensional (1D) arrays but two-dimensional (2D) arrays. The 1D FSSP is described as follows: given a one-dimensional array of $n$ identical cellular automata, including a {\it general} at one end that is activated at time $t = 0$, we want to design the automata such that, \textit{at some future time}, all the cells will \textit{simultaneously} and, \textit{for the first time}, enter a special {\it firing} state. 

Some questions may arise:
 
\begin{itemize}
\item
How can we synchronize multi-dimensional arrays?  
\item
Can we expand those 2D FSSP algorithms proposed so far to 3D arrays, or more generally to multi-dimensional arrays? 
\item 
Can we generalize those 2D FSSP algorithms to generalized ones, in which an initial general is located at any position of the array?
\item
What is a lower bound of time steps needed for synchronizing multi-dimensional arrays with a general at one corner?
\item
What is a lower bound for the generalized case?
\end{itemize}

In the present paper, we attempt to answer these questions by proposing a new, simple recursive-halving based synchronization algorithm for 2D rectangle cellular automata. The algorithm can synchronize any 2D rectangle array of size $m \times n$ with a general at one corner in $m + n + \max(m, n)-3 $ steps. An implementation in terms of local transition rules is also given on a 2D cellular automaton, not only for the array with a general at one corner but a generalized case. 

The algorithms proposed in this paper are interesting in the following view points.
\begin{itemize}
\item
The 2D algorithm proposed is isotropic with respect to shape of a given rectangle array, i.e. no need to control the FSSP algorithm for longer-than-wide and wider-than-long input rectangles.
\item
The algorithm proposed can be easily expanded to 3D arrays, even to multi-dimensional arrays.
\item
The algorithm is a natural generalization of the well-known 1D FSSP algorithms developed by Waksman [1966], Balzer [1967] and Gerken [1987]. It gives us a new view point of those classical 1D FSSP algorithms based on recursive-halving.
\item
The algorithm can be expanded to a generalized FSSP solution, where an initial general is at an arbitrary position of a given array.
\item
The algorithm can be generalized to an optimum-time generalized FSSP solution.
\end{itemize}

In Section 2 we give a description of the 2D FSSP and review some basic results on 2D FSSP algorithms. Section 3 defines the recursive-halving marking on 1D arrays and gives some preliminary lemmas for the construction of 2D FSSP algorithms. In Sections 4, 5, and 6 we present a new 2D FSSP algorithm based on the recursive-halving marking and several multi-dimensional expansions.  Two implementations in terms of 2D cellular automata are also presented for the optimum-time FSSP algorithms.  Most of the descriptions of the multi-dimensional FSSP algorithms are based on the 2D FSSP algorithms.  Some expanded and generalized theorems for multi-dimensional arrays are given without proofs.

\section{Firing Squad Synchronization Problem on Two-Dimensional Arrays}

Figure 1 shows a finite two-dimensional (2D) cellular automaton consisting of $m \times n $ cells. Each cell is an identical (except the border cells) finite-state automaton. The array operates in lock-step mode in such a way that the next state of each cell (except border cells) is determined by both its own present state and the present states of its north, south, east and west neighbors.  All cells ({\it soldiers}), except the north-west corner cell ({\it general}), are initially in the quiescent state at time $t = 0$ with the property that the next state of a quiescent cell with quiescent neighbors is
the quiescent state again. At time $t = 0$, the north-west corner cell $\mathrm{C}_{1 1}$ is in the {\it fire-when-ready} state, which is the initiation signal for the array. The firing squad synchronization problem is to determine a description (state set and next-state function) for cells that ensures all cells enter the {\it fire} state at exactly the same time and for the first time. 
The tricky part of the problem is that the same kind of soldier having a fixed number of states must be synchronized, regardless of the size $m \times n$ of the array. The set of states and next state function must be independent of $m$ and $n$. 

\begin{figure}[h]
\begin{center}
\includegraphics[width=7cm, clip] {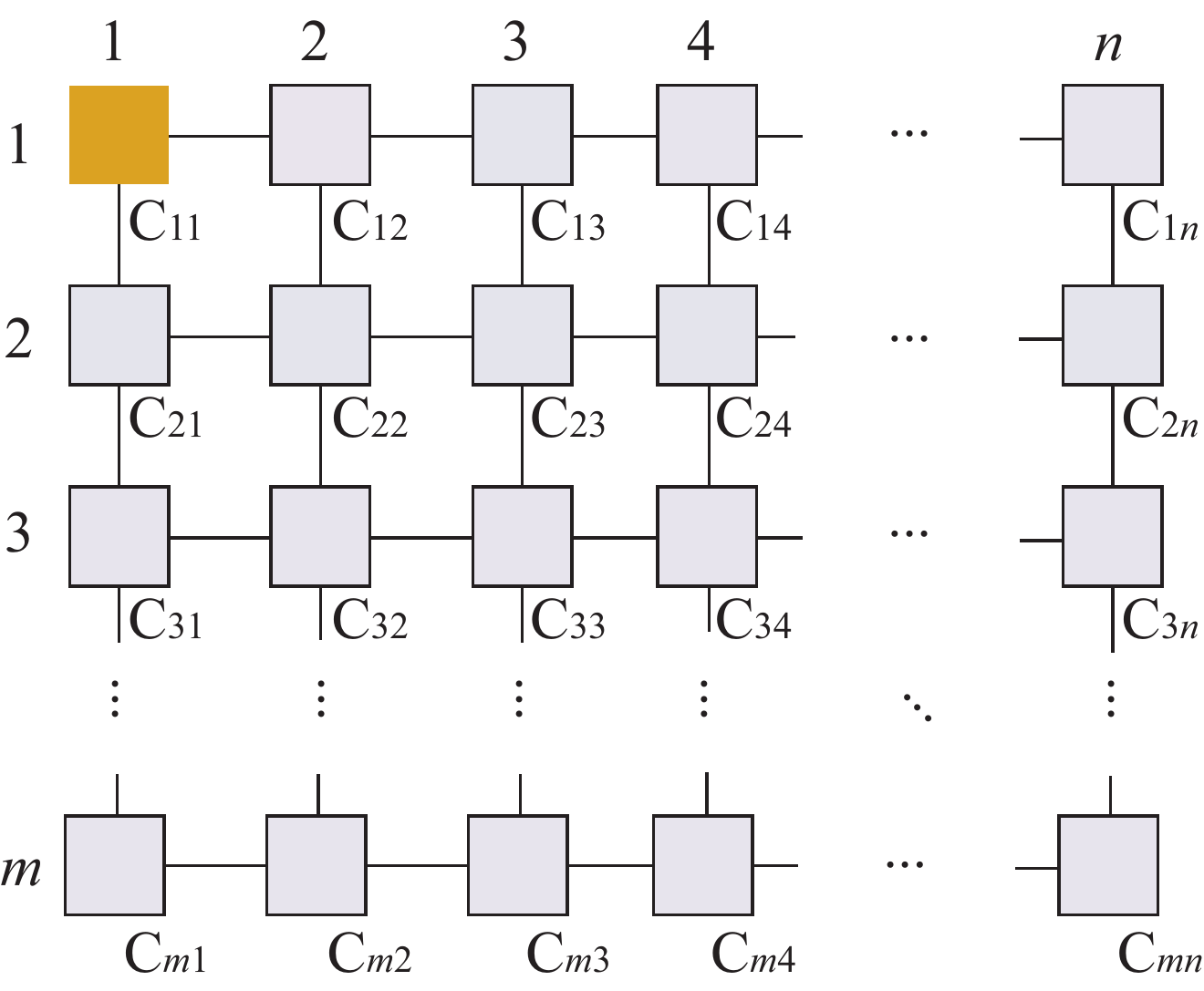}
\caption[]{\label{Fig1} A two-dimensional (2D) cellular automaton.}
\end{center}
\end{figure}

  The problem was first solved by J. McCarthy and M. Minsky who presented a $3n$-step algorithm for 1D cellular array of length $n$.
In 1962, the first optimum-time, i.e. $(2n-2)$-step, synchronization algorithm was presented by Goto [1962], with each cell having several thousands of states. 
Waksman [1966] presented a 16-state optimum-time synchronization algorithm.  Afterward, Balzer [1967] and Gerken [1987] developed an eight-state algorithm and a seven-state synchronization algorithm, respectively, thus decreasing the number of states required for the synchronization.
Mazoyer [1987] developed a six-state synchronization algorithm which, at present, is the algorithm having the fewest states for 1D arrays.

On the other hand, several synchronization algorithms on 2D arrays have been proposed by Beyer [1969], Grasselli [1975], Shinahr [1974], Szwerinski [1982],  Schmid [2003], Schmid and Worsch [2004], Umeo, Maeda, Hisaoka and Teraoka [2006], and Umeo and Uchino [2008].
It has been shown independently by Beyer [1969] and Shinahr [1974] that there exists no 2D cellular automaton that can synchronize any 2D array of size $m \times n$ in less than $m + n + \max(m, n)- 3$ steps. In addition they first proposed an optimum-time synchronization algorithm that can synchronize any 2D array of size $m \times n$ in optimum $m + n + \max(m, n)- 3$ steps. Shinahr [1974] gave a 28-state implementation.  Umeo, Hisaoka and Akiguchi [2005] presented a new 12-state synchronization algorithm operating in optimum-step, realizing a smallest solution to the rectangle synchronization problem at present.   

As for the time optimality of the 2D FSSP algorithms, the following theorems have been shown.

\vspace{2mm}
\noindent \textbf{Theorem 1}$^{{\rm Beyer}  \ [1969], \ {\rm Shinahr} \ [1974]}$
There exists no cellular automaton that can synchronize any 2D array of size $m \times n$ in less than $m + n + \max(m, n)- 3$ steps, where the general is located at one corner of the array. 
\vspace{2mm}

\vspace{2mm}
\noindent \textbf{Theorem 2}$^{{\rm Shinahr} \ [1974],\ {\rm Umeo, \ Hisaoka, \ and \ Akiguchi \ [2005]}}$
There exists a cellular automaton that can synchronize any 2D array of size $m \times n$ at exactly $m + n + \max(m, n)- 3$ steps, where the general is located at one corner of the array. 
\vspace{2mm}

\section{Recursive-Halving Marking}

In this section, we develop a marking schema for 1D arrays referred to as \textit{recursive-halving marking}.  The marking schema prints a special mark on cells in a cellular space defined by the recursive-halving marking.  The marking itself is based on a 1D FSSP synchronization algorithm.  It will be effectively used for constructing multi-dimensional FSSP algorithms operating in optimum-time.

 Let $S$ be a 1D cellular space consisting of cells C$_{i}$, C$_{i+1}$, ..., C$_{j}$, denoted by [$i$...$j$], where $j > i$.  Let $|S|$ denote the number of cells in $S$, that is $|S|= j-i+1$.  A center cell(s) C$_{x}$ of $S$ is defined by\\
\begin{equation}
x=\begin{cases}
         (i+j)/2 & \text{$|S|$:  odd} \\
         (i+j-1)/2, (i+j+1)/2& \text{$|S|$: even.}
   \end{cases}  
\end{equation}

The recursive-halving marking for a given cellular space $S$ = [1...$n$] is defined as follows: 

\begin{center}
Recursive-Halving Marking: RHM \hrulefill
{\small
\begin{quote}
\baselineskip=8pt
\textbf{Algorithm RHM($S$)}\\
\textbf{begin}\\
\hspace{5mm}\textbf{if} $|S| \geq 2$ \textbf{then}\\
\hspace{5mm}\hspace{5mm}	\textbf{if} $|S|$ is odd \textbf{then}\\
\hspace{5mm}\hspace{10mm} 	\textbf{mark} a center cell C$_{x}$ in $S$;\\
\hspace{5mm}\hspace{10mm}		$S_{L}$:= [$1$...$x$]; $S_{R}$:= [$x$...$n$];\\
\hspace{5mm}\hspace{10mm}	\textbf{RHM$_{\rm L}$($S_{L}$)};	
				\textbf{RHM$_{\rm R}$($S_{R}$)};\\
\hspace{5mm}\hspace{5mm}	\textbf{else}\\
\hspace{5mm}\hspace{10mm}	     \textbf{mark} center cells C$_{x}$ and C$_{x+1}$ in $S$;\\
\hspace{5mm}\hspace{10mm}	         $S_{L}$:= [$1$...$x$]; $S_{R}$:= [$x+1$...$n$];\\
\hspace{5mm}\hspace{10mm}	\textbf{RHM$_{\rm L}$($S_{L}$)};
				\textbf{RHM$_{\rm R}$($S_{R}$)};\\
\textbf{end}\\
\end{quote}
}
\hrulefill
\end{center}

\begin{center}
Left-Side Recursive-Halving Marking: RHM$_{\rm L}$ \hrulefill
{\small
\begin{quote}
\baselineskip=8pt
\textbf{Algorithm RHM$_{\rm L}$($S$)}\\
\textbf{begin}\\
\hspace{5mm}\textbf{while} $|S| > 2$ \textbf{do}\\
\hspace{5mm}\hspace{5mm}	\textbf{if} $|S|$ is odd \textbf{then}\\
\hspace{5mm}\hspace{10mm} 	\textbf{mark} a center cell C$_{x}$ in $S$;\\
\hspace{5mm}\hspace{10mm}		$S_{L}$:= [$1$...$x$]; \textbf{RHM$_{\rm L}$($S_{L}$)};\\ 
\hspace{5mm}\hspace{5mm}	\textbf{else}\\
\hspace{5mm}\hspace{10mm}	     \textbf{mark} center cells C$_{x}$ and C$_{x+1}$ in $S$;\\
\hspace{5mm}\hspace{10mm}	         $S_{L}$:= [$1$...$x$]; \textbf{RHM$_{\rm L}$($S_{L}$)};\\
\textbf{end}\\
\end{quote}
}
\hrulefill
\end{center}

\vspace{10mm}

\begin{center}
Right-Side Recursive-Halving Marking: RHM$_{\rm R}$ \hrulefill
{\small
\begin{quote}
\baselineskip=8pt
\textbf{Algorithm RHM$_{\rm R}$($S$)}\\
\textbf{begin}\\
\hspace{5mm}\textbf{while} $|S| > 2$ \textbf{do}\\
\hspace{5mm}\hspace{5mm}	\textbf{if} $|S|$ is odd \textbf{then}\\
\hspace{5mm}\hspace{10mm} 	\textbf{mark} a center cell C$_{x}$ in $S$;\\
\hspace{5mm}\hspace{10mm}		$S_{R}$:= [$x$...$n$]; \textbf{RHM$_{\rm R}$($S_{R}$)};\\
\hspace{5mm}\hspace{5mm}	\textbf{else}\\
\hspace{5mm}\hspace{10mm}	     \textbf{mark} center cells C$_{x}$ and C$_{x+1}$ in $S$;\\
\hspace{5mm}\hspace{10mm}	         $S_{R}$:= [$x+1$...$n$]; \textbf{RHM$_{\rm R}$($S_{R}$)};\\
\textbf{end}\\
\end{quote}
}
\hrulefill
\end{center}

For example, we consider a cellular space $S$ = [1...$15$] consisting of 15 cells. The first center cell is C$_{8}$, then the second one is  C$_{4}$, C$_{5}$ and C$_{11}$, C$_{12}$, and the last one is C$_{2}$, C$_{3}$, C$_{13}$, C$_{14}$, respectively. In case $S$ = [1...$17$], we get C$_{9}$, C$_{5}$, C$_{13}$, C$_{3}$, C$_{15}$, and C$_{2}$, C$_{16}$ after four iterations. 

\begin{figure}[h]
\begin{center}
\includegraphics[width=\textwidth, clip] {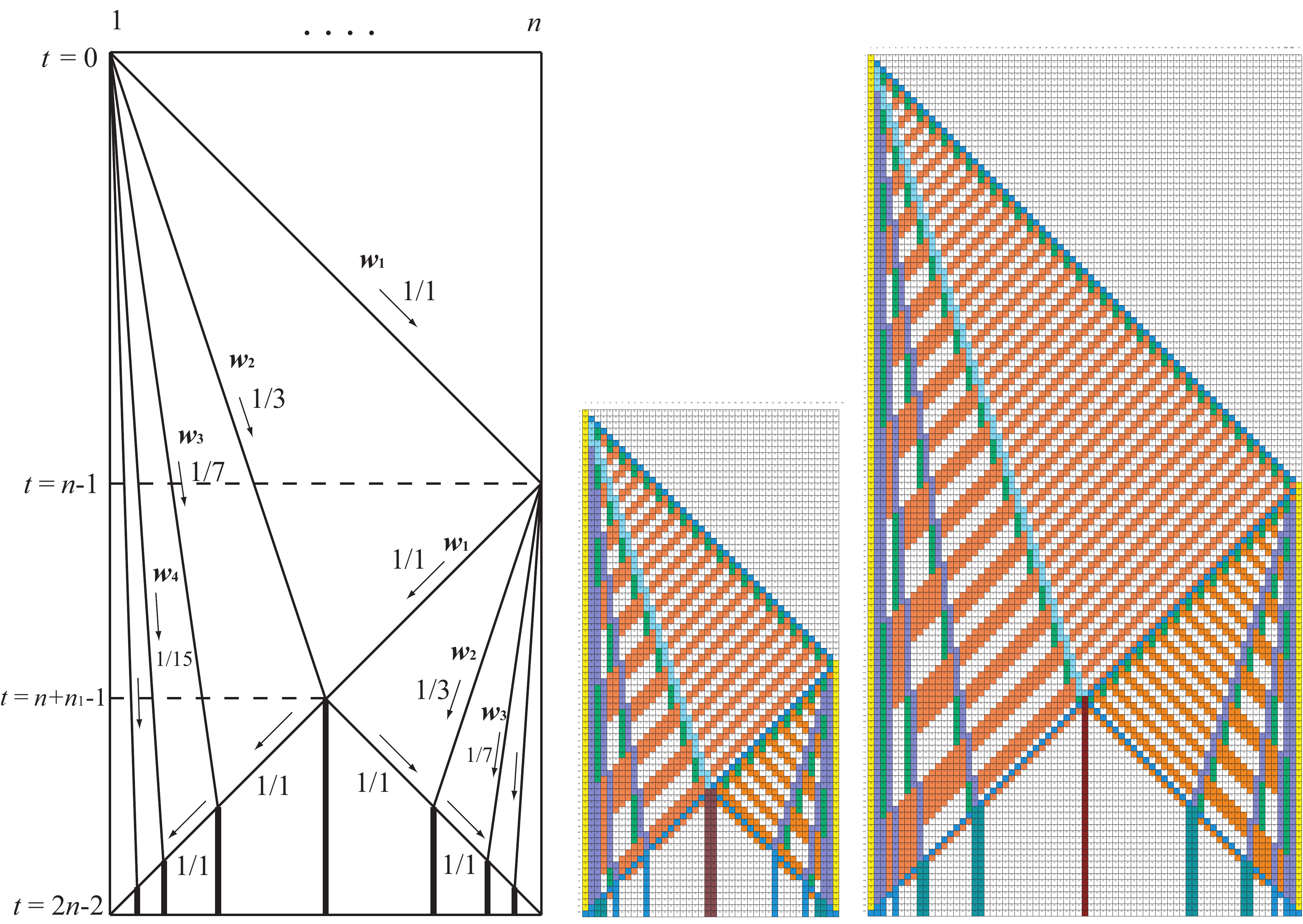}
\caption[]{\label{Fig2} Space-time diagram for recursive-halving marking on 1D array of length $n$ (left) and some snapshots for the marking on 42 (middle) and 71 (right) cells, respectively.}
\end{center}
\end{figure}

 Figure \ref{Fig2} (left) shows a space-time diagram for the marking. At time $t=0$, the leftmost cell C$_{1}$ generates an infinite set of signals $w_{1}, w_{2}, ..., w_{k}$, .., each propagating in the right direction at $1/(2^{k}-1)$ speed, where $k=1, 2, 3, ...,$ . 
 The $1/1$-speed signal $w_{1}$ arrives at C$_{n}$ at time $t=n-1$.  Then, the rightmost cell C$_{n}$ also emits an infinite set of signals $w_{1}, w_{2}, ..., w_{k}$, .., each propagating in the left direction at $1/(2^{k}-1)$ speed, where $k=1, 2, 3, ...,$ . The readers can find that each crossing of two signals, shown in Fig. 2 (left), enables the marking at middle points defined by the recursive-halving. A finite state realization for generating the infinite set of signals above is a well-known technique employed in Balzer [1967], Gerken [1987], and Waksman [1966] for the implementations of the optimum-time synchronization algorithms on 1D arrays.

We have developed a simple implementation of the recursive-halving marking on a 13-state, 314-rule cellular automaton. In Fig. 2 (middle and right) we present several snapshots for the marking on 42 and 71 cells, respectively. 
Thus we have:

\vspace{2mm}
\noindent \textbf{Lemma 3}
There exists a 1D 13-state, 314-rule cellular automaton that can print the recursive-halving marking in any cellular space of length $n$ in $2n-2$ steps.
\vspace{2mm}

An optimum-time complexity $2n-2$ needed for synchronizing cellular space of length $n$ in the classical WBG-type (Waksman [1966], Balzer [1967], and Gerken [1987]) FSSP algorithms can be interpreted as follows: 
Let $S$ be a cellular space of length $n=2n_{1}+1$, where $n_{1} \geq 1$. The first center mark in $S$ is printed on cell C$_{n_{1}+1}$ at time $t_{1{\rm D-center}}=3n_{1}$. Additional $n_{1}$ steps are required for the markings thereafter, yielding a final synchronization at time $t_{1{\rm D-opt}}=3n_{1}+n_{1}= 4n_{1}= 2n-2$.
In the case $n=2n_{1}$, where $n_{1} \geq 1$, the first center mark is printed  simultaneously on cells C$_{n_{1}}$ and C$_{n_{1}+1}$ at time $t_{1{\rm D-center}}=3n_{1}-1$. 
Additional $n_{1}-1$ steps are required for the marking and synchronization thereafter, yielding the final synchronization at time $t_{1{\rm D-opt}}=3n_{1}-1+n_{1}-1= 4n_{1}-2= 2n-2$.
\begin{equation}
t_{{\rm 1D-center}}=\begin{cases}
         3n_{1} & \text{$|S|=2n_{1}+1$,} \\
         3n_{1}-1& \text{$|S|=2n_{1}$.}
   \end{cases}  
\end{equation}

Thus, additional $t_{{\rm 1D-sync}}$ steps are required for the synchronization for a cellular space with the recursive-halving marks:
\begin{equation}
t_{{\rm 1D-sync}}=\begin{cases}
         n_{1} & \text{$|S|=2n_{1}+1$,} \\
         n_{1}-1& \text{$|S|=2n_{1}$.}
   \end{cases}  
\end{equation}

\begin{figure}[h]
\begin{center}
\includegraphics[width=\textwidth, clip] {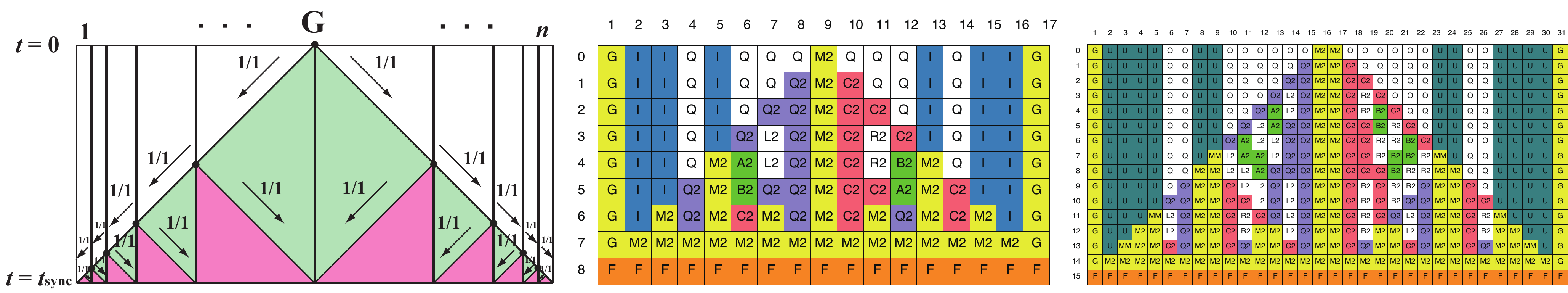}
\caption[]{\label{Fig3} Space-time diagram for synchronizing a cellular space with recursive-halving marking (left) and some snapshots for the synchronization on 17 (middle) and 32 (right) cells, respectively.}
\end{center}
\end{figure}

In this way, it can be easily seen that any cellular space of length $n$ with the recursive-halving marking initially with a general on a center cell or two generals on adjacent center cells can be synchronized in $\lceil n/2 \rceil-1$ optimum-steps.  In Fig. \ref{Fig3}, we illustrate a space-time diagram for synchronizing a cellular space with recursive-halving marking (left) and some snapshots for the synchronization on 17 (middle) and 32 (right) cells, respectively. Thus we have: 

\vspace{2mm}

\noindent \textbf{Lemma 4}
Any 1D cellular space $S$ of length $n$ with the recursive-halving marking initially with a general(s) on a center cell(s) in $S$ can be synchronized in $\lceil n/2 \rceil-1$ optimum-steps.

\vspace{2mm}

As was seen, the first marking of center cell(s) plays an important role. We print a special mark for the first center cell(s) of a given cellular space. On the other hand, for the center cells generated thereafter are marked with a different symbol from the first one.  

\begin{figure}[h]
\begin{center}
\includegraphics[width=12cm, clip] {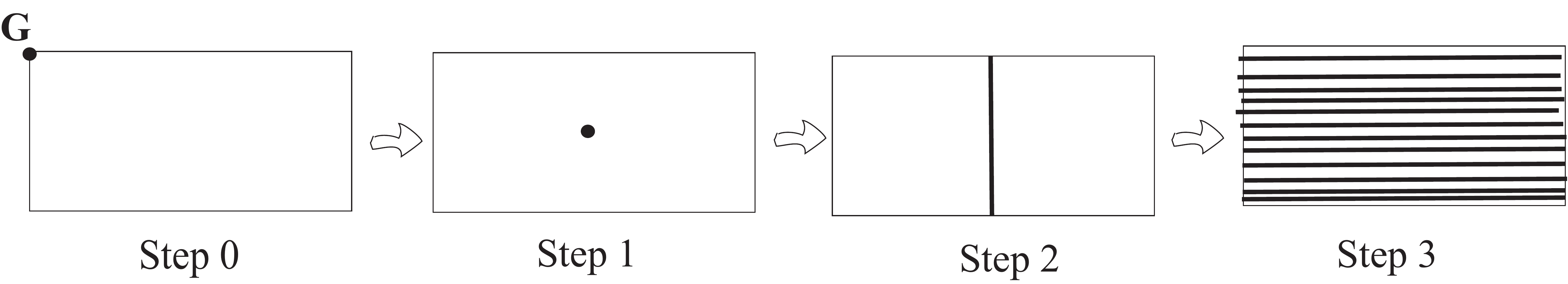}
\caption[]{\label{Fig-FSSP-2D} Synchronization schema for 2D cellular automaton.}
\end{center}
\end{figure}

\section{An Optimum-Time 2D FSSP Algorithm $\mathcal{A}_{1}$}

\subsection{Overview of the Algorithm $\mathcal{A}_{1}$}
We assume that an initial general G is on the north-west corner cell C$_{1 1}$
 of a given array of size $m \times n$. The algorithm consists of three phases: a marking phase, a pre-synchronization phase and a final synchronization phase.
An overview of the 2D synchronization algorithm $\mathcal{A}_{1}$ is as follows:  
\vspace{2mm}

\noindent \textbf{Step 1.}  \textbf{Start} the recursive-halving marking for cells on each row and column, \textbf{find} a \textit{center cell(s) of the given array}, and \textbf{generate} a new general(s) on the center cell(s).  Note that a crossing(s) of the center column(s) with the center row(s) is a center cell(s) of the array.
 
\noindent \textbf{Step 2.} \textbf{Pre-synchronize} the center column(s) using Lemma 4, which is initiated by the general in Step 1. Every cell on the center column(s) acts as a general at the following Step 3.

\noindent \textbf{Step 3.} \textbf{Synchronize} each row using Lemma 4, initiated by the general generated in Step 2.  This yields the final synchronization of the array.

\vspace{2mm}

\begin{figure}[h]
\begin{center}
\includegraphics[width=10cm, clip] {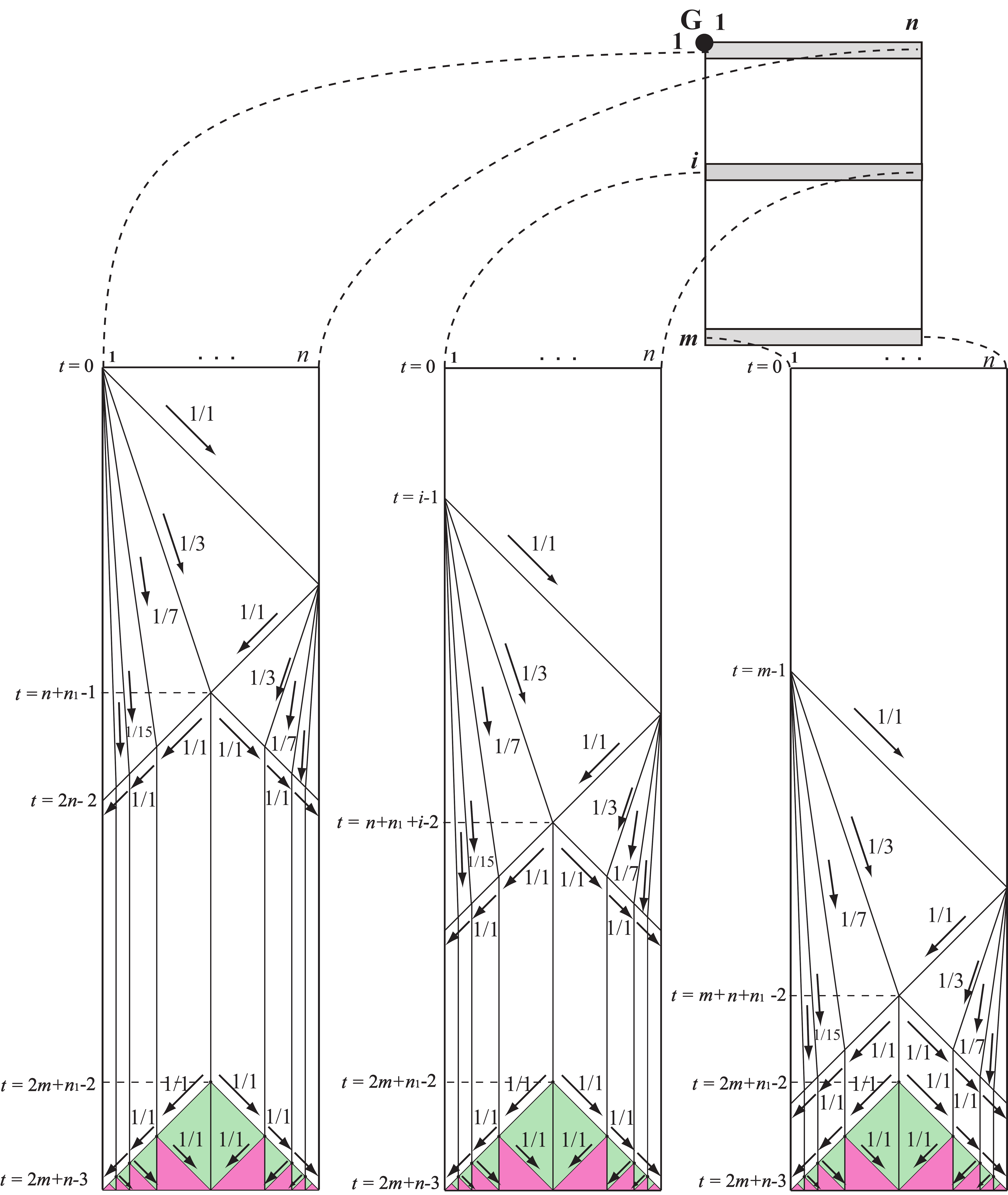}
\caption[]{\label{Fig7-1} Space-time diagrams for the synchronization algorithm on the 1st, $i$th, and $m$th rows of a longer-than-wide rectangle array of size $m \times n$, respectively.}
\end{center}
\end{figure}

Figure \ref{Fig-FSSP-2D} illustrates the synchronization schema for 2D cellular automaton.  We assume that $m= 2m_{1}+1, n= 2n_{1}+1$, where $m_{1}, n_{1} \geq 1$. The algorithm operates as follows: 

\begin{enumerate}
\item
At time $t=0$ an initial general on the north-west corner emits a $1/1$-speed signal along the first row and column to print recursive-halving marks.  Once a center mark is printed, it is copied to the adjacent row and column.
At time $t=3m_{1}+n_{1}$, a center mark of the center column of the array is marked, and the center mark of the center row is marked at time $t=3n_{1}+m_{1}$.
The center of the array is marked at time $t= t_{2{\rm D-center}}=\max(3m_{1}+n_{1}, 3n_{1}+m_{1})$.

\item
Using Lemma 4, the center column will be synchronized with a tentative pre-firing state at time $t= t_{2{\rm D-center}} + m_{1}$.  

\item
Once the center column could be synchronized with the pre-firing state, then the cell C$_{i, n_{1}+1}$ initiates the synchronization for the $i$th row for each $i$ such that $1 \leq i \leq m$.  Using Lemma 4, for any $i, 1 \leq i \leq m$, the $i$th row will be synchronized at time $t=  t_{2{\rm D-center}} + m_{1}+ n_{1} = \max(3m_{1}+n_{1}, 3n_{1}+m_{1}) + m_{1}+ n_{1}= \max(2m_{1}+1, 2n_{1}+1) + 2m_{1}+ 2n_{1}-1 = m+n+\max(m, n)-3$. Thus, the array can be synchronized at time $t=m+n+\max(m, n)-3$ in optimum-steps.

\end{enumerate}

\begin{figure}[h]
\begin{center}
\includegraphics[width=12cm, clip] {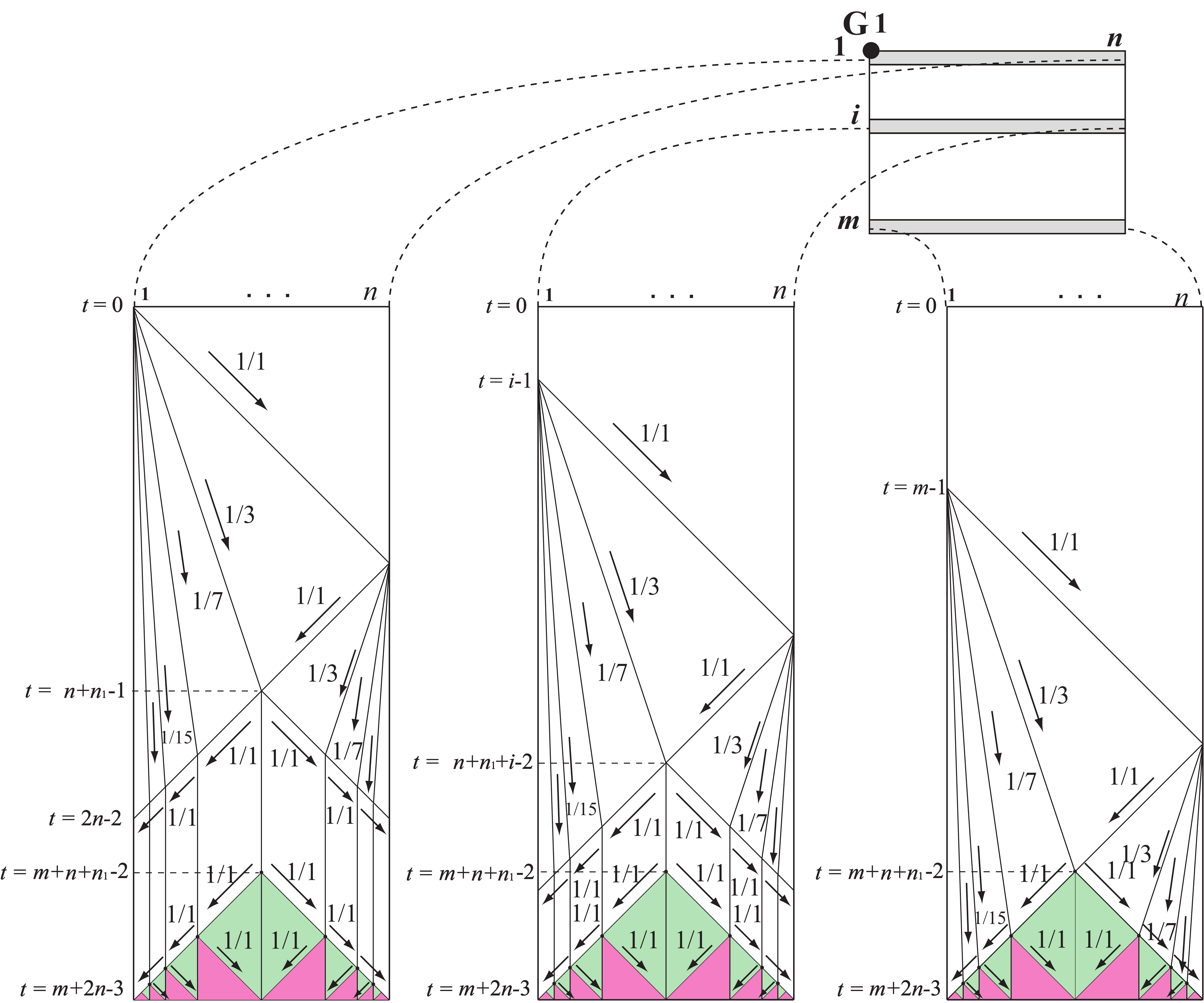}
\caption[]{\label{Fig7-2} Space-diagrams of the synchronization algorithm on the 1st, $i$th, and $m$th rows of a wider-than-long rectangle array of size $m \times n$, respectively.}
\end{center}
\end{figure}

\begin{figure}[h]
\begin{center}
\includegraphics[width=15cm, clip] {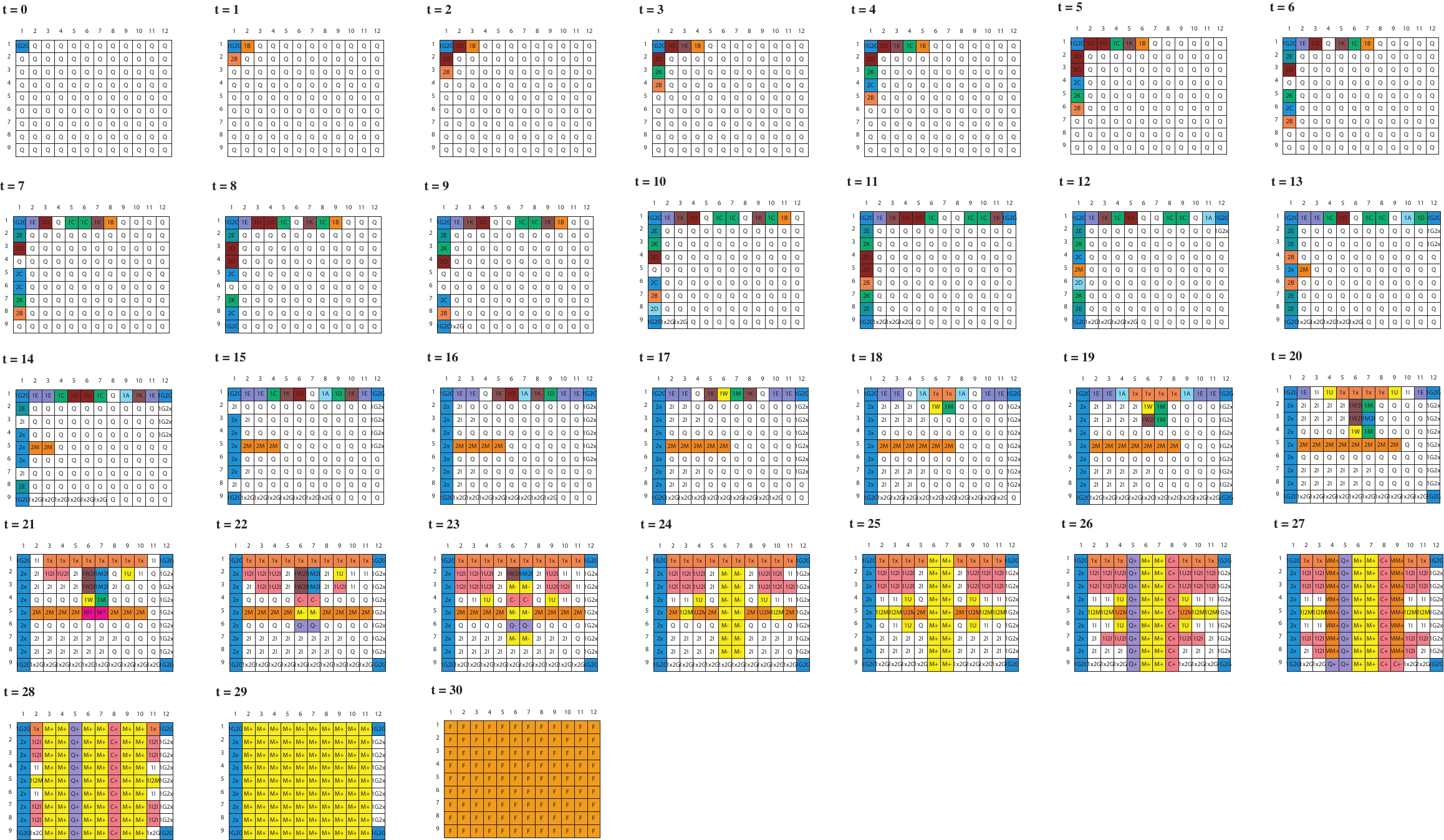}
\caption[]{\label{9x12} Snapshots for synchronization on a $9 \times 12$ array.}\end{center}
\end{figure}

\begin{figure}[h]
\begin{center}
\includegraphics[width=15cm, clip] {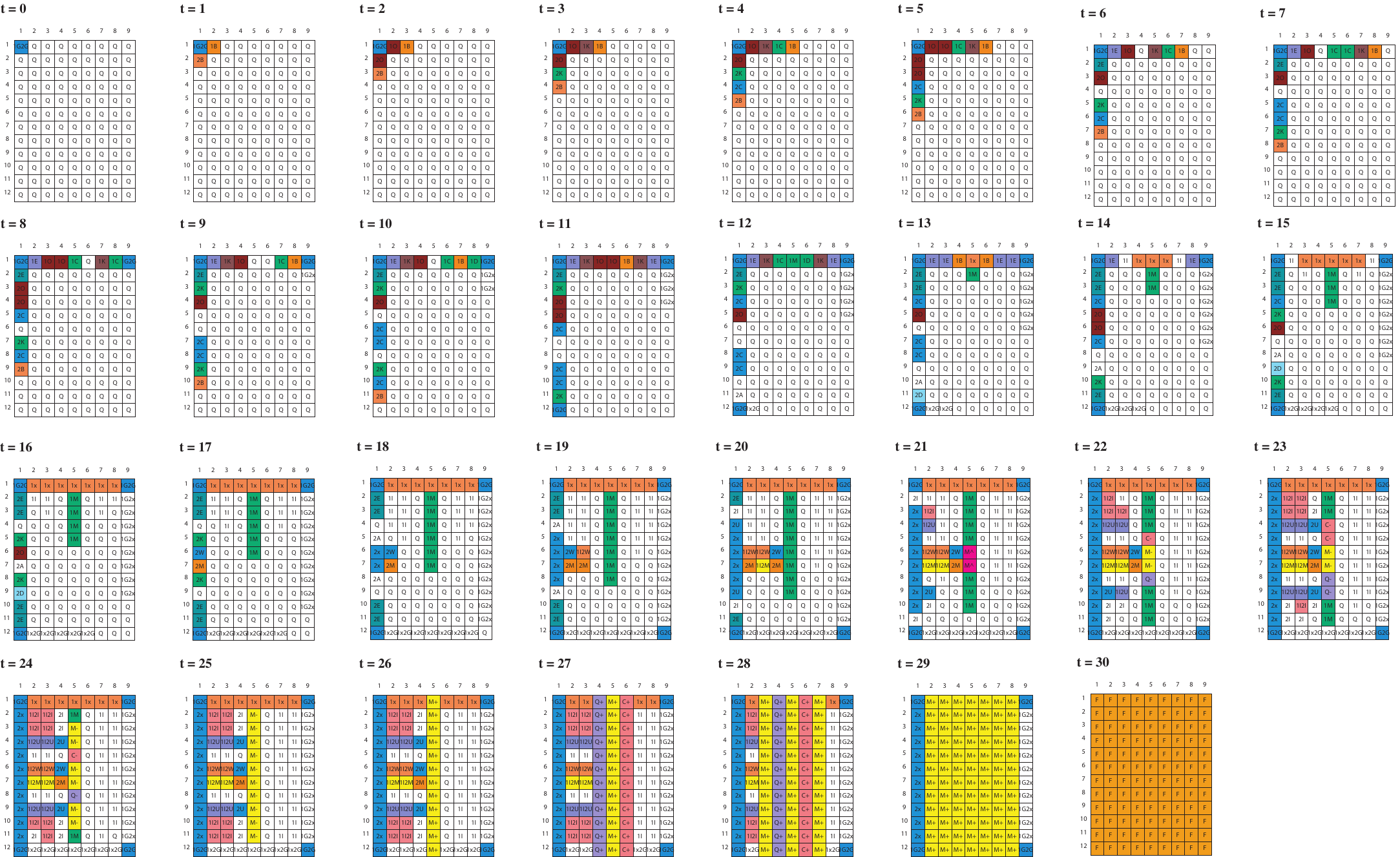}
\caption[]{\label{12x9} Snapshots for synchronization on a $12 \times 9$ array.}
\end{center}
\end{figure}

Note that the signal propagation for the recursive-halving marking and the wake-up signal for the synchronization are made by the same 1/1 speed.  Thus, the synchronization can be performed successfully for each column and the array can be synchronized in optimum steps. 
Figures \ref{Fig7-1} and \ref{Fig7-2} illustrate a space-time diagram for the recursive-halving marking and the synchronization operations on the 1st, $i$th, and $m$th row of a longer-than-wide and wider-than-long array, respectively.  One can see that each marking operation has been finished before the arrival of the first wake-up signal for the synchronization. The algorithm operates  in optimum-steps in a similar way for the rectangles such as case 1: $m= 2m_{1}+1, n= 2n_{1}$, case 2: $m= 2m_{1}, n= 2n_{1}+1$, and case 3: $m= 2m_{1}, n= 2n_{1}$.

Thus, we can establish the following theorem.

\vspace{2mm}
\noindent \textbf{Theorem 5}
 The synchronization algorithm $\mathcal{A}_{1}$  can synchronize any $m \times n$ rectangular array in optimum $m + n + \max(m, n)- 3$ steps. 
\vspace{2mm}

We have implemented the algorithm $\mathcal{A}_{1}$ on a 2D cellular automaton having 60 states and 13633 local rules.  In Figures \ref{9x12} and \ref{12x9} we present some snapshots of the synchronization processes of the algorithm on $9 \times 12$ and $12 \times 9$ arrays, respectively. 

\section{Expansion to Multi-Dimensional Arrays}

\subsection{Three-Dimensional Arrays}

In this section, we show that there exists no algorithm that can synchronize any 3D array of size $m \times n \times \ell$ with a general at an arbitrary corner in less than $m+n+ \ell + max(m, n, \ell)-4$ steps.

\vspace{2mm}
\noindent {\bf Theorem 6}
The minimum time in which the firing squad synchronization could occur is no earlier than $m+n+ \ell + max(m, n, \ell)-4$ for any 3D array of size $m \times n \times \ell$ with a general at an arbitrary corner cell.

\textit{Proof}. The proof is made by contradiction.  Without loss of generality, we assume that $\ell \leq m \leq n$. It is assumed that there is a cellular automaton $\mathcal{M}$ that can synchronize an array of size $m_{0} \times n_{0} \times \ell_{0}$ $(\ell_{0} \leq m_{0} \leq n_{0})$ at step $t=t_{0}$ such that:

\begin{center}
\begin{equation}
t_{0} < m_{0} +2n_{0} + \ell_{0} -4
\end{equation}
\end{center}

Now consider the state of cell C$_{m_{0} \ 1 \ \ell_{0}}$ at time $t=t_{0}$.  Let $i$ and $k$ be any integer such that $1 \leq i \leq m_{0}$ and $1 \leq k \leq \ell_{0}$. Consider the signal propagation from the cell C$_{1\ 1\ 1}$ to C$_{m_{0} \ 1 \ \ell_{0}}$ via any cell C$_{i \ n_{0} \ k}$.  It takes:
\begin{center}
\begin{eqnarray}
(i-1)+ (n_{0}-1)+(k-1)+(m_{0}-i)+(n_{0}-1)+ (\ell_{0}-k) \nonumber\\ 
=m_{0}+2n_{0}+\ell_{0}-4
\end{eqnarray}
\end{center}
steps for the signal to travel from C$_{1\ 1\ 1}$ to C$_{m_{0} \ 1 \ \ell_{0}}$ via any cell C$_{i \ n_{0} \ k}$.
The state of the cell C$_{m_{0} \ 1 \ \ell_{0}}$ at step $t=t_{0}$ entered the final firing state unaffected by any cells on the plane \{C$_{i \ n_{0} \ k}| 1 \leq i \leq m_{0},  1 \leq k \leq \ell_{0}$ \}.  Therefore, if another three-dimensional array of size $m_{0} \times n_{0} \times \ell_{0}$ was added to the right side of the original array (that is, the new array is of size $m_{0} \times 2n_{0} \times \ell_{0}$), the cell C$_{m_{0} \ 1 \ \ell_{0}}$ would still enter the final firing state at step $t=t_{0}$.
This is because the cell structure $\mathcal{M}$ is fixed, cell operation is deterministic and nothing has changed as far as the cell C$_{m_{0}\ 1\ \ell_{0}}$ is concerned.  Since $t_{0} < m_{0} +2n_{0} + \ell_{0} -4$, the cell C$_{m_{0} \ 1 \ \ell_{0}}$ will still be in a quiescent state at time $t=t_{0}$.
Therefore  the cell structure does not represent a solution and this is a contradiction.  In a similar way, the argument carries over in the cases such as $\ell \leq n \leq m, n \leq \ell \leq m$, ..., and so forth.  {\flushright $\square$}

\begin{figure}[h]
\begin{center}
\includegraphics[width=12cm] {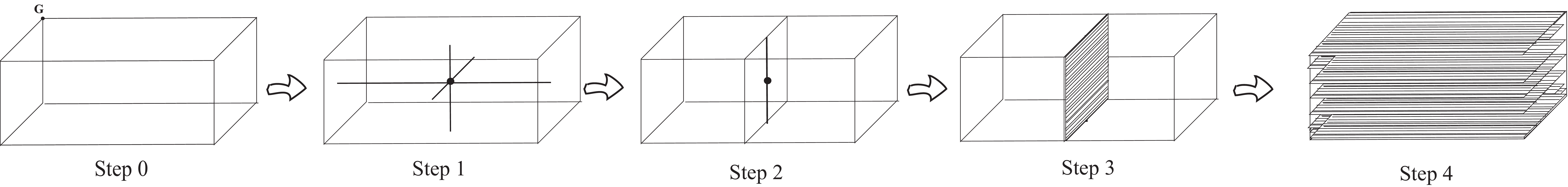}
\caption[]{\label{Fig-FSSP-3D} Synchronization scheme for a three-dimensional cellular automaton.}
\end{center}
\end{figure}

The synchronization algorithm $\mathcal{A}_{1}$ for 2D arrays can be easily expanded to 3D arrays. See Figure \ref{Fig-FSSP-3D} which illustrates the synchronization schema for 3D cellular automaton.
Thus, we have:

\vspace{2mm}
\noindent \textbf{Theorem 7}
 There exists an optimum-time synchronization algorithm $\mathcal{A}_{2}$ that can synchronize any three-dimensional array of size $n_{1} \times n_{2} \times n_{3}$ with a general at C$_{1, 1, 1}$ in optimum $n_{1} + n_{2} + n_{3} + \max(n_{1}, n_{2}, n_{3})- 4$ steps. 
\vspace{2mm}

\subsection{Multi-Dimensional Arrays}

A $k$D FSSP algorithm $\mathcal{A}_{3}$ is sketched as follows: 
 
\vspace{2mm}
\noindent \textbf{Step 1.}  \textbf{Start} the recursive-halving marking on cells along each dimension, \textbf{find} a \textit{center cell(s) of the given array}, \textbf{generate} a general(s) on the center cell(s), and \textbf{pre-synchronize} the center point(s): zero-dimensional sub-array of the array. \\ 
\noindent \textbf{Step 2.} \textbf{Pre-synchronize} a 1D sub-array along the 1st dimension containing the pre-synchronized center cell.\\
\noindent \textbf{Step 3. - Step $k$.} For $j=2$ to $k-1$, by increasing the number of dimensions, \textbf{pre-synchronize} a $j$D sub-array containing the pre-synchronized $(j-1)$D sub-array.\\
\noindent \textbf{Step $k+1$.} \textbf{Synchronize} the $k$D array. This yields the final synchronization of the given array.\\
\vspace{2mm}

Theorems 6 and 7 can be expanded to the $k$D arrays.

\vspace{2mm}
\noindent \textbf{Theorem 9}
There exists no cellular automaton that can synchronize any $k$D array of size $n_{1} \times n_{2} \times ... \times n_{k}$ with a general at C$_{1, 1, ..., 1}$  in less than $\sum^{k}_{i=1}n_{i} + \max(n_{1}, n_{2}, ..., n_{k})- k-1$ steps

\vspace{2mm}

\noindent \textbf{Theorem 10}
 There exists an optimum-time synchronization algorithm $\mathcal{A}_{3}$ that can synchronize any $k$D array of size $n_{1} \times n_{2} \times ... \times n_{k}$ with a general at C$_{1, 1, ..., 1}$ in optimum $\sum^{k}_{i=1}n_{i} + \max(n_{1}, n_{2}, ..., n_{k})- k-1$ steps. 

\vspace{2mm}

\begin{figure}[h]
\begin{center}
\includegraphics[width=6cm, clip] {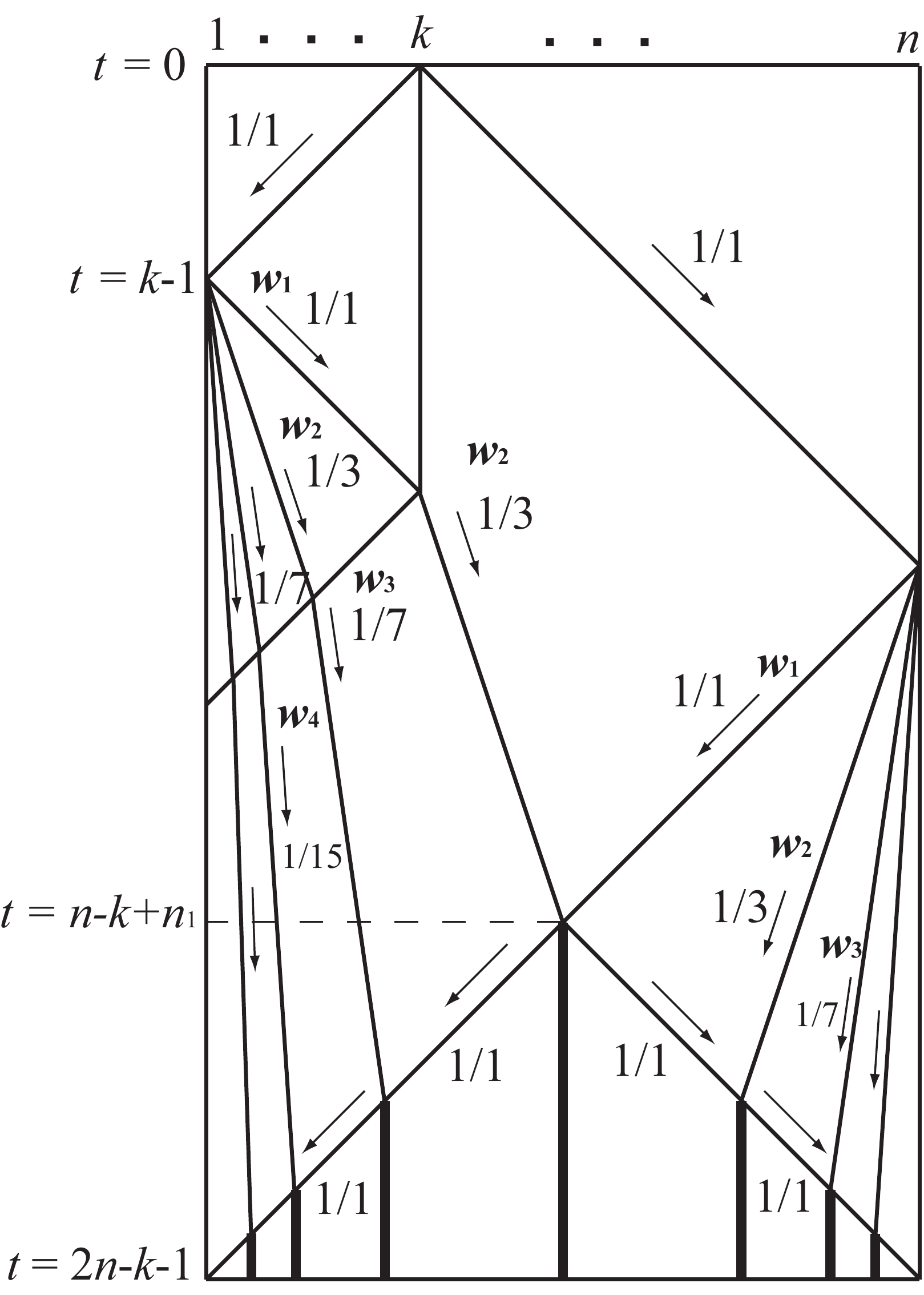}
\caption[]{\label{Fig2-7-2} Space-time diagram for recursive-halving marking on 1D array of length $n$ with a general at any position.}
\end{center}
\end{figure}

\section{Generalization as to General's Position}

\subsection{Generalized FSSP on 1D Arrays}

The recursive-halving marking scheme on 1D array can be easily expanded to the generalized case where the initial general is located at any position of the array. Figure \ref{Fig2-7-2} illustrates a space-time diagram for the recursive-halving marking on 1D array of length $n$ with a general on C$_{k}, 1 \leq k \leq n$. The marking is based on the generalized FSSP algorithm proposed by Moore and Langdon [1968].

We have seen the following theorems for the 1D generalized case.

\vspace{2mm}
\noindent \textbf{Theorem 11} $^{{\rm Moore \ and \ Langdon} \ [1968]}$ The minimum time in which the generalized firing squad synchronization could occur is $n -2 + max(k, n-k+1)$ steps, where the general is located on the $k$th cell from left end. 

\vspace{2mm}

\noindent \textbf{Theorem 12} $^{{\rm Moore \ and \ Langdon} \ [1968]}$ There exists a 17-state cellular automaton that can synchronize any one-dimensional array of length $n$ in optimum $n -2 + max(k, n-k+1)$ steps, where the general is located on the $k$th cell from left end. 

\vspace{2mm}

An optimum-time complexity $n -2 + max(k, n-k+1)$ needed for synchronizing cellular space of length $n$ in the classical ML-type (Moore and Langdon [1968]) generalized FSSP algorithm can be interpreted as follows: 
Let $S$ be a cellular space of length $n=2n_{1}+1$, where $n_{1} \geq 1$. The first center mark in $S$ is printed on cell C$_{n_{1}+1}$ at time $t_{1{\rm Dg-center}}=3n_{1}-\min(k-1, n-k)$. Additional $n_{1}$ steps are required for the markings thereafter, yielding a final synchronization at time $t_{1{\rm Dg-opt}}=3n_{1}-\min(k-1, n-k)+n_{1}= 4n_{1}-\min(k-1, n-k)= 2n-2-\min(k-1, n-k)=n-2+\max(k, n-k+1)$.

In the case $n=2n_{1}$, where $n_{1} \geq 1$, the first center mark is printed  simultaneously on cells C$_{n_{1}}$ and C$_{n_{1}+1}$ at time $t_{1{\rm Dg-center}}=3n_{1}-1-\min(k-1, n-k)$. Note that two cells C$_{n_{1}}$ and C$_{n_{1}+1}$ are pre-synchronized at time $t_{1{\rm Dg-center}}=3n_{1}-1-\min(k-1, n-k)$. 
Additional $n_{1}-1$ steps are required for the marking thereafter, yielding the final synchronization at time $t_{1{\rm Dg-opt}}=3n_{1}-1-\min(k-1, n-k)+n_{1}-1= 4n_{1}-2-\min(k-1, n-k)= 2n-2-\min(k-1, n-k)=n-2+\max(k, n-k+1)$.

\begin{equation}
t_{{\rm 1Dg-center}}=\begin{cases}
         3n_{1}-\min(k-1, n-k) & \text{$|S|=2n_{1}+1$,} \\
         3n_{1}-1-\min(k-1, n-k)& \text{$|S|=2n_{1}$.}
   \end{cases}  
\end{equation}

In addition, $t_{{\rm 1Dg-sync}}$ steps are required for the synchronization for a cellular space with the recursive-halving marks:

\begin{equation}
t_{{\rm 1Dg-sync}}=\begin{cases}
         n_{1} & \text{$|S|=2n_{1}+1$,} \\
         n_{1}-1& \text{$|S|=2n_{1}$.}
   \end{cases}  
\end{equation}

\subsection{Generalized FSSP on 2D Arrays}

By a similar method employed in Section 4, we can develop the following theorem for the generalized case. 

\vspace{2mm}
\noindent \textbf{Theorem 13}
There exists no 2D cellular automaton that can synchronize any 2D array of size $m \times n$ with an initial general on C$_{r, s}$ in less than $m + n + \max(m, n)-\min(r, m-r+1)-\min(s, n-s+1)-1$ steps, where $1 \leq r \leq m, 1 \leq s \leq n$. 
\vspace{2mm}

Now we are going to present a generalized optimum-time FSSP Algorithm $\mathcal{A}_{4}$ for 2D arrays. We assume that an initial general G is on the cell C$_{r, s}$ of a given array of size $m \times n$, where $1 \leq r \leq m, 1 \leq s \leq n$. The algorithm consists of three phases: a marking phase, a pre-synchronization phase and a final synchronization phase.
An overview of the 2-D synchronization algorithm $\mathcal{A}_{4}$ is as follows:
  
\vspace{2mm}
\noindent \textbf{Step 1.}  \textbf{Start} the recursive-halving marking for cells on each row and column, \textbf{find} a \textit{center cell(s) of the given array}, and \textbf{generate} a new general(s) on the center cell(s).  Note that a crossing(s) of the center column(s) with the center row(s) is a center cell(s) of the array.
 
\noindent \textbf{Step 2.} \textbf{Pre-synchronize} the center column(s) using Lemma 4, which is initiated by the general in step 1. Every cell on the center column(s) acts as a general at the next Step 3.

\noindent \textbf{Step 3.} \textbf{Synchronize} each row using Lemma 4, initiated by the general generated in Step 2.  This yields the final synchronization of the array.
\vspace{2mm}

The array can be synchronized at time $t=m + n + \max(m, n)-\min(r, m-r+1)-\min(s, n-s+1)-1$ in optimum-steps.

\vspace{2mm}
\noindent \textbf{Theorem 14}
 There exists an optimum-time synchronization algorithm $\mathcal{A}_{4}$ that can synchronize any $m \times n$ rectangular array  with a general at C$_{r, s}$ in optimum $m + n + \max(m, n)-\min(r, m-r+1)-\min(s, n-s+1)-1$ steps, where $1 \leq r \leq m, 1 \leq s \leq n$. 
\vspace{2mm}

\begin{figure}[h]
\begin{center}
\includegraphics[width=15cm, clip] {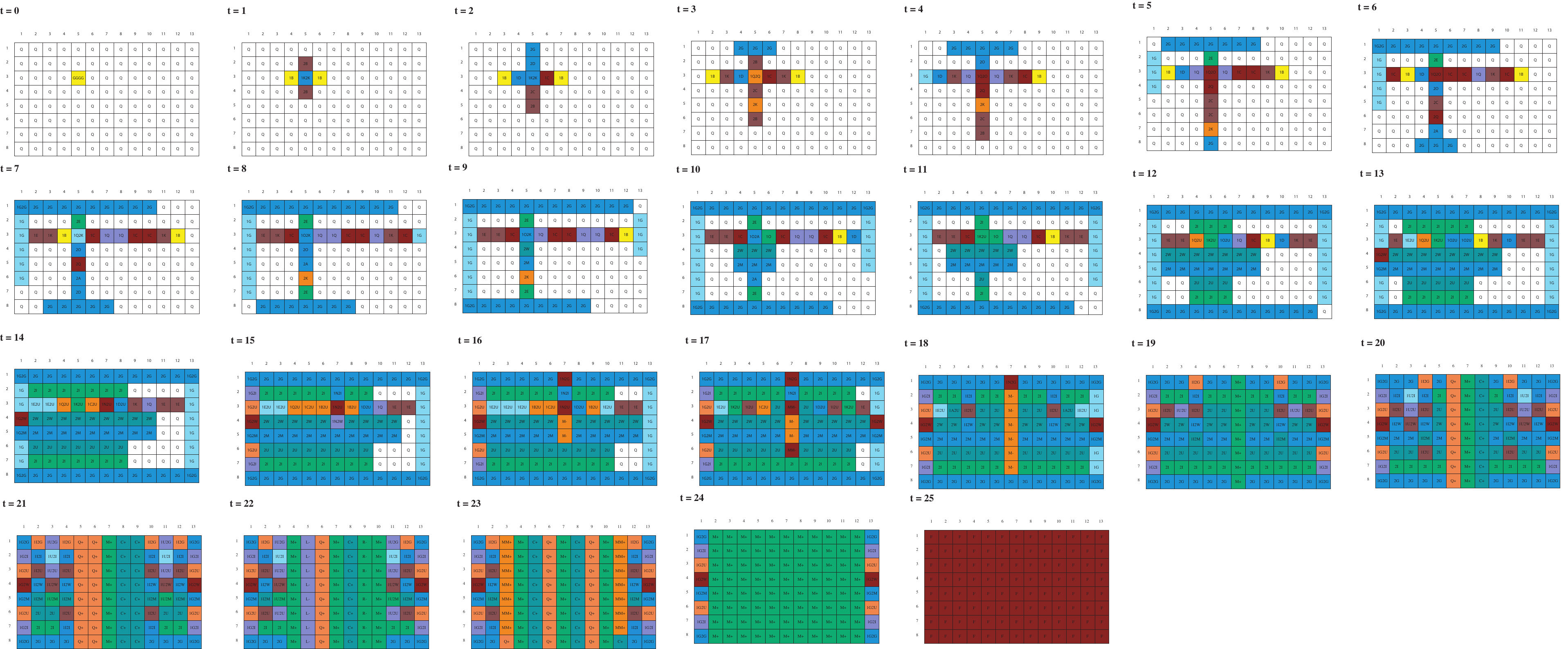}
\caption[]{\label{g-simulation-8x13} Snapshots of the generalized synchronization algorithm $\mathcal{A}_{4}$ on an $8 \times 13$ array with a general on C$_{3, 5}$.}
\end{center}
\end{figure}

We have implemented the algorithm $\mathcal{A}_{4}$ on a 2D cellular automaton having 269 states and 163662 local rules. We have checked the rule set for any array of size $m \times n$, with $2 \leq m, n \leq 100$, and any general's position in the array. In Figure \ref{g-simulation-8x13} we present some snapshots of the synchronization processes of the algorithm on an $8 \times 13$ array with a general on C$_{3, 5}$.

\subsection{Generalized FSSP on Multi-Dimensional Arrays}

Theorems 13 and 14 can be expanded to three or more dimensional arrays.

\vspace{2mm}

\noindent {\bf Theorem 15}
The minimum time in which the firing squad synchronization could occur is no earlier than $n_{1}+n_{2}+ n_{3} + \max(n_{1}, n_{2}, n_{3})-\min(r_{1}, n_{1}-r_{1}+1)-\min(r_{2}, n_{2}-r_{2}+1)- \min(r_{3}, n_{3}-r_{3}+1)-1$ for any 3D array of size $n_{1} \times n_{2} \times n_{3}$ with a general at C$_{r_{1}, r_{2}, r_{3}}$.

\vspace{2mm}
\noindent \textbf{Theorem 16}
 There exists an optimum-time synchronization algorithm $\mathcal{A}_{5}$ that can synchronize any 3D array of size $n_{1} \times n_{2} \times n_{3}$ with a general at C$_{r_{1}, r_{2}, r_{3}}$ in optimum $n_{1} + n_{2} + n_{3} + \max(n_{1}, n_{2}, n_{3})- \min(r_{1}, n_{1}-r_{1}+1)-\min(r_{2}, n_{2}-r_{2}+1)- \min(r_{3}, n_{3}-r_{3}+1)-1$ steps.

\vspace{2mm}

\noindent \textbf{Theorem 17}
There exists no cellular automaton that can synchronize any $k$D array of size $n_{1} \times n_{2} \times ... \times n_{k}$ with a general at C$_{r_{1}, r_{2}, ..., r_{k}}$  in less than $\sum^{k}_{i=1}n_{i} +\max(n_{1}, n_{2}, ..., n_{k})-\sum^{k}_{i=1}\min(r_{i}, n_{i}-r_{i} +1)-1$ steps.

\vspace{2mm}
\noindent \textbf{Theorem 18}
 There exists an optimum-time synchronization algorithm $\mathcal{A}_{6}$ that can synchronize any $k$D array of size $n_{1} \times n_{2} \times ... \times n_{k}$ with a general at  C$_{r_{1}, r_{2}, ..., r_{k}}$ in optimum $\sum^{k}_{i=1}n_{i} +\max(n_{1}, n_{2}, ..., n_{k})-\sum^{k}_{i=1}\min(r_{i}, n_{i}-r_{i} +1)-1$ steps. 

\vspace{2mm}

\section{Conclusions}

We have proposed a new class of optimum-time multi-dimensional FSSP algorithms based on recursive-halving marking.  The class includes the well-known optimum-time FSSP algorithms developed by Waksman [1964], Balzer [1966] and Gerken [1987] with a general at one end and Moore and Langdon [1968] with a general at any position. 

\vspace{2mm}
\noindent \textbf{Acknowledgments} 
The authors would like to thank reviewers for helpful comments.

\end{document}